
\magnification=1200
\baselineskip=20pt
\centerline{\bf PERTURBATIONS OF A TOPOLOGICAL DEFECT AS}
\centerline{\bf  A THEORY OF COUPLED SCALAR FIELDS
IN CURVED SPACE}
\centerline{\bf INTERACTING WITH AN EXTERNAL VECTOR POTENTIAL}
\vskip1pc
\centerline {\bf Jemal Guven }
\vskip1pc
\it
\centerline {Instituto de Ciencias Nucleares }
\centerline {Universidad Nacional Aut\'onoma de M\'exico}
\centerline {A. Postal 70-543. 04510 M\'exico, D. F., MEXICO} \rm
\centerline{(guven@unamvm1.bitnet)}
\vskip1pc
\centerline{\bf Abstract}
\vskip1pc
{\leftskip=1.5cm\rightskip=1.5cm\smallskip\noindent
The evolution of small irregularities in a
topological defect which propagates
on a curved background spacetime is examined.
These are described by a system of
coupled scalar wave equations on the worldsheet of the
unperturbed defect which is not only
manifestly covariant under world sheet diffeomorphisms but
also under local normal frame rotations. The scalars couple
both through the surface torsion of the background
worldsheet geometry which acts as a vector
potential and through an effective mass
matrix which is a sum of a quadratic in the extrinsic curvature
and a linear term in the spacetime curvature.
The coupling simplifies enormously for many physically interesting
geometries. This introduces a framework
for examining the stability of topological defects
generalizing both our earlier work on of the perturbations
of domain walls and the work of Garriga and Vilenkin on
perturbations about a class of spherically symmetric defects in
de Sitter space. \smallskip}

\vskip2pc
\noindent{\bf I. INTRODUCTION}
\vskip1pc
Topological defects of one form or another
are expected to appear as by-products of
phase transitions that occured
in the early universe. Their cosmological
implications, however, appear to
depend sensitively on their stability with respect to perturbations.
For example, an instability in the
geometry of a closed cosmic string could disrupt its collapse to
form a black hole.[1] Recently, Garriga and Vilenkin
undertook an examination of the stability of spherically symmetric
topological defects nucleating
in de Sitter space.[2] The approximation they use
is to model the defect as a membrane propagating on a
curved background spacetime.

In this paper, we examine the evolution of small irregularities on
a topological defect moving in a general
curved background spacetime in the same approximation without any
restrictions on the symmetry of the defect.

In an earlier paper, we treated
the perturbations of domain walls.[3]
The relevant covariant measure of the perturbation then
is its projection onto the
normal to the worldsheet. We were able to  show that this
scalar satisfies a Klein-Gordon equation
on the geometry of the unperturbed worldsheet,
coupling in a universal manner through an effective mass
both to the world-sheet scalar curvature and the
traced projection of the spacetime Ricci curvature onto the world-sheet.
This provided a generalization of
the wave equation derived in Ref.[2]
describing perturbations of domain walls in Minkowski space.

In the case of a lower dimensional defect there will be one scalar
corresponding to the projection of the perturbation in the world-sheet
onto each normal direction. A new geometrical
structure which is anti-symmetric in its normal indices
also appears. The geometrical role it plays in
perturbation theory is that of a vector potential
ensuring that the scalar field equations transform
covariantly under local normal frame rotations.
These scalars will generally satisfy
a system of wave equations which are
coupled not only through an effective mass matrix but also through this
vector potential. In particular this introduces a
derivative coupling between the fields.

We begin in Section II with a
derivation of the exact equations of motion for the defect.
Our approach to perturbation theory in Section III
will be to expand the action describing the evolution of the
defect in a manifestly covariant way out to second
order in the perturbation about a given classical solution.
We find that the easiest way to evaluate the
second variation is to develop a formalism which is manifestly
covariant from the beginning. We model this on
Hawking and Ellis's treatment of the second variation of the arc-length
about a geodesic curve.[4]
Geodesics, however, can be a poor guide
to the behavior of higher dimensional surfaces.
The proper length along a curve has no higher dimensional analog;
the curvature of a connection has no one-dimensional analog.
It is therefore extremely gratifying that
the formal expression one obtains is strikingly similar to the
geodesic result when the parameterization along the latter is
not affine. We exploit the classical theory of surfaces
to bridge the gap between formal mathematics and a tractable
system of equations with which
one can begin to do physics.[5]

In Section IV. we discuss the equations of motion decribing
perturbations on various background geometries.
In practice, one is interested in perturbations
about defects possessing some level of symmetry.
It is then, of course,
sufficient to develop perturbation theory in a manner which is
tailored to the symmetry. In Ref.[2], doing just this, it was shown that on
a spherically symmetric string of maximum radius in de Sitter space
these equations not only decouple completely but each component
tends to mimic the single scalar characterizing the perturbation on a
domain wall in a de Sitter space of one lower dimension.
It is probably fair to say,
however, that in the absence of a more general framework to steer by
one is at a loss to provide an entirely
adequate interpretation of the physics.
It is not clear what features of the
underlying geometry are responsible for the
simplification in perturbation theory discovered in Ref.[2].
Do we always expect the effective mass to be tachyonic?
We attempt to provide sufficient criteria
determining when the equations will decouple.
In particular, we demonstrate that whenever the
world-sheet of the defect can be embedded as a hypersurface
in some lower dimensional geometry and the codimension of the
world-sheet is one or two, the equations of motion completely decouple.
As a special case we rederive Eq.(58) of Ref.[2].

\vskip2pc
\noindent{\bf II. THE EQUATIONS OF MOTION}
\vskip1pc

Let us consider an oriented surface $m$ of dimension $D$
described by the timelike surface
$$x^\mu=X^\mu(\xi^a)\,,\eqno(2.1)$$
$\mu=0,\cdots, N-1$, $a=0,\cdots, D-1$,
embedded in an $N$-dimensional spacetime $M$
described by the metric $g_{\mu\nu}$. The $D$ vectors
$$e_a =X^\mu_{,a}\partial_\mu$$
form a basis of tangent vectors to $m$ at each point of $m$.
The metric induced on the world sheet is then
given by
$$\gamma_{ab}= X^\mu_{,a}
X^\nu_{,b}\,g_{\mu\nu} = g(e_a,e_b)\,.\eqno(2.2)$$
The action which describes
the dynamics of our system is the most simple generally covariant
action one can associate with the
surface, proportional to the area swept out by
the world sheet of the surface as it evolves:

$$S[X^\mu,X^\mu_{,a}]=-\sigma\int_m d^{D}\xi \sqrt{-\gamma}
\,,\eqno(2.3)$$
The constant of proportionality $\sigma$ represents the constant
(positive) energy density in the surface in its rest  frame.
If the area is infinite the associated action will be infinite.
However, the change in area corresponding to a variation in the
embedding of compact support will always be finite.

We confine our attention to
closed defects without physical boundaries. The only boundary
of the world-sheet is then the spacelike boundary, $\partial m_t$,
we introduce to implement the variational
principle, marking the temporal limits of the
world-sheet on which the initial and
final configurations are fixed.

The equations of motion of the defect are given by the
extrema of $S$ subject to variations
$X^\mu(\xi)\to X^\mu(\xi)+\delta X^\mu(\xi)$ which vanish on
$\partial m_t$:

$$-{\delta S\over \delta X^\mu} \equiv
\sigma\left[\Delta X^\mu + \Gamma^\mu_{\alpha\beta}(X^\nu)
\gamma^{ab} X^\alpha_{,a} X^\beta_{,b}\right]=0\,,\eqno(2.4)$$
where $\Delta$ is the scalar Laplacian
$$\Delta ={1\over\sqrt{\gamma}}
\partial_a(\sqrt{\gamma}\gamma^{ab}\partial_b)\,,$$
and
$\Gamma^\mu_{\alpha\beta}$ are the spacetime Christoffel symbols
evaluated on $m$. We return to the derivation at the end of this
section. Eq.(2.4) is clearly a higher dimensional generalization of the
geodesic equation describing the motion of a point defect. Even in
Minkowski space, however, this equation is highly non-linear.
The feature of string theory which makes it
tractable is the fact that the
world-sheet is two-dimensional and any two dimensional
metric is conformally flat.[6]

Despite the nice analogy,
this form of the equations of motion is not very
useful in practice.  This is because all but $N-D$ linear combinations
of these equations are identically satisfied.
To see this, we note that, both on shell and off, Gauss's equation
(see Eq.(4.8a) below) can be rewritten in the form
$$\nabla_b X^\mu_{,a} + \Gamma^\mu_{\alpha\beta}(X^\nu)
X^\alpha_{,a} X^\beta_{,b}=K_{ab}^{(i)} n^{(i)\,\mu}
\,,$$
where $\nabla_a$ is the world-sheet covariant derivative
compatible with $\gamma_{ab}$,
$n^{(i)}$ is the $i^{\rm th}$ unit normal to the worldsheet, $i=1,\cdots,
N-D$, and the corresponding $i^{\rm th}$
extrinsic curvature tensor $K_{ab}^{(i)}$ is defined by [5]
(we introduce the notation ${\cal D}_a= X^\mu_{,a} D_\mu$)
$$K^{(i)} _{ab}=-X^\mu_{,a} X^\nu_{,b} D_\nu n^{(i)}_{\mu} =
-g(e_a, {\cal D}_b n^{(i)})\,.\eqno(2.5)$$
The tangential projections
of the Euler-Lagrange derivatives of $S$ therefore vanish identically:
$$ {\delta S\over \delta X^\mu} X^\mu_{,b}=0\,.\eqno(2.6)$$
As we have remarked in Ref.[3], the geometrical reason for this redundancy
is the invariance of the action with respect to
world sheet  diffeomorphisms.

It is now clear that the
equations describing the world sheet are entirely equivalent to the
$N-D$ equations
$$K^{(i)}=0\,.\eqno(2.7)$$
These are just the equations describing an extremal
surface and are well known in the
mathematical literature.[7] They provide an obvious generalization of the
more familiar notion of extremal hypersurface.

To derive Eq.(2.4), we note that
$$\delta S =
-{1\over 2}\sigma\int_m d^{D}\xi \sqrt{-\gamma}\gamma^{ab}
{\cal D}_\delta g(e_a,e_b)\,,\eqno(2.8)$$
where we introduce the spacetime vector field
$$\delta = \delta X^\mu\partial_\mu\,,\eqno(2.9)$$
to characterize the deformation in the world-sheet.
We also set ${\cal D}_\delta =\delta X^\mu D_\mu$.
The key observation in the derivation is that
the Lie derivative of the vector field
$\delta$ with respect to $e_a$ vanishes
(the proof is sketched in chapter 4 of [4]):
$${\cal D}_\delta e_a = {\cal D}_a \delta\,.\eqno(2.10)$$
Now
$$\eqalign{\gamma^{ab}{\cal D}_\delta g(e_a,e_b)=& 2 \gamma^{ab}g({\cal
D}_\delta e_a, e_b)\cr
		          = & 2 \gamma^{ab} g({\cal D}_a \delta, e_b)\cr
                  =  & 2 \gamma^{ab} \left[{\cal D}_a
g(\delta, e_b)- g(\delta, {\cal D}_a e_b)\right]\,,\cr}$$
The first term on the last line can be reorganized as follows
$$\sqrt{-\gamma}\gamma^{ab} {\cal D}_a g(\delta, e_b)=
{\cal D}_a\left[\sqrt{-\gamma}\gamma^{ab}  g(\delta, e_b)\right]-
\nabla_a (\sqrt{-\gamma}\gamma^{ab} ) g(\delta, e_b)$$
to extract a divergence. Because  $\delta$ vanishes on
$\partial m_t$ this term will also vanish there.
We are left with the simple formal expression
$$\delta S = \sigma \int d^D\xi \,
g(\delta, {\cal D}_a\left[\sqrt{-\gamma}\gamma^{ab} e_b\right])\,.\eqno(2.11)$$
The equations of motion are therefore
$${\cal D}_a\left[\sqrt{-\gamma}\gamma^{ab} e_b\right]=0\,,\eqno(2.12)$$
the spacetime components of which reproduce Eq.(2.4).

\vskip2pc
\noindent{\bf III. THE QUADRATIC ACTION}
\vskip1pc

At lowest order, the dynamics of the irregularities
in the defect is still expected to be accurately
described by the action  Eq.(2.3).

The approach we will follow will be to expand the
action out to quadratic order about the
classical solution satisfying Eq.(2.7).
When this is done, it will be a relatively straightforward
matter to write down the corresponding equations of motion.

As we have seen, variations along tangential directions
correspond to world-sheet diffeomorphisms.
We can, however, provide a diffeomorphism invariant
description of the perturbation $\delta X^\mu$ in the wall by
constructing the $N-D$ scalars
$$\Phi^{(i)} \equiv n^{(i)}_\mu\delta X^\mu\eqno(3.1)$$
representing the independent projections of the
spacetime vector $\delta X^\mu$ onto the different normal directions.
The choice of the $\Phi^{(i)}$ is not unique reflecting the fact that
the defining relations for the normal vectors

$$g(e_a, n^{(i)})=0\,,\quad g(n^{(i)},n^{(j)})=\delta^{(i)(j)}$$
only determine these vectors up to
$N-D$ dimensional frame rotations. If the geometry possesses some
symmetry, it is very convenient to choose the normal vectors
so that they reflect the symmetry.

In our earlier treatment of domain walls it was possible to exploit
Gaussian normal coordinates based on the
the world-sheet hypersurface to facilitate the calculation.
In general, there is no simple analog of
Gaussian coordinates corresponding to a lower dimensional embedding.[8]
It is fortunate, therefore, that the covariant formalism we
have been pursuing is tractable.

We now evaluate the second variation of the action
at its stationary points. This is given by

$$\delta^2 S = \sigma \int d^D\xi\,\, g(\delta, {\cal D}_\delta {\cal D}_a
\left[\sqrt{-\gamma}\gamma^{ab} e_b\right])\,.
\eqno(3.2)$$
Thus, the relevant equation is
$${\cal D}_\delta {\cal D}_a\left[\sqrt{-\gamma}\gamma^{ab} e_b\right]=0
\,.\eqno(3.3)$$
While formally Eq.(3.3) describes small
perturbations, it is not very useful in its present form.
What we need to do is to cast Eq.(3.3) explicitly as a linear system of
coupled scalar wave equations
$${\cal L}_{(i)(j)}\Phi^{(j)}=0\,,\eqno(3.4)$$
Our task is to find the linear hyperbolic partial
differential operator, ${\cal L}$.
Such an equation can be derived from an action of the form

$$S={1\over2}\delta^2 S=
{1\over 2}\int d^D\xi\sqrt{-\gamma} \Phi^{(i)}\tilde
{\cal L}_{(i)(j)}\Phi^{(j)}\,,$$
where $\tilde {\cal L}_{(i)(j)}$ is some other
linear hyperbolic operator.
As we will see $\tilde {\cal L}_{(i)(j)}$ is linear in first derivatives
of the fields $\Phi^{(i)}$. As a consequence
it will not coincide with ${\cal L}_{(i)(j)}$.
We are always free to
symmetrize $\tilde {\cal L}_{(i)(j)}$ (but not
${\cal L}_{(i)(j)}$) with respect to the indices $i$ and $j$.

Let us examine the projection of the LHS of Eq.(3.3) on $n^{(i)}$
$$g(n^{(i)}, {\cal D}_\delta {\cal D}_a\left[\sqrt{-\gamma}\gamma^{ab}
e_b\right])=0\,.$$
We need only consider vector fields $\delta$ which are normal to $m$.
The idea is to push ${\cal D}_\delta$ to the right
through ${\cal D}_a$ and $e_a$.
Let us begin then by exploiting the spacetime
Ricci identity on the spacetime
vector field $v=\sqrt{-\gamma}\gamma^{ab} e_b $

$${\cal D}_\delta {\cal D}_a v =
{\cal D}_a{\cal D}_\delta v+ ^N\! R(e_a,\delta) v\,.\eqno(3.5)$$
We note that
by exploiting the projection tensor,

$$h^{\mu\nu}= \gamma^{ab} X^\mu_{,a}X^{\nu}_{,b}=
g^{\mu\nu}- n^{(k)\,\mu} n^{(k)\,\nu}\,,\eqno(3.6)$$
we can express
$$\gamma^{ab}{}g(\delta,  ^N\! R(e_a,\delta)e_b) =
\left[^N\! R_{\mu\nu} n^{(i)\,\mu} n^{(j)\,\nu}
- ^N\! R_{\mu\alpha\nu\beta}\, n^{(i)\,\mu} n^{(k)\,\alpha} n^{(j)\,\nu}
n^{(k)\,\beta}\right]\Phi^{(i)}\Phi^{(j)}\,.$$
We now decompose ${\cal D}_a{\cal D}_\delta v$ into three terms as follows:
$$\eqalign{g(n^{(i)}, {\cal D}_a{\cal D}_\delta \left[\sqrt{-\gamma}\gamma^{ab}
e_b\right])
= &{\cal D}_a \left[\sqrt{-\gamma}\gamma^{ab} g(n^{(i)}, {\cal D}_\delta
e_b)\right] \cr
-&{\cal D}_\delta (\sqrt{-\gamma}\gamma^{ab})  g({\cal D}_a n^{(i)}, e_b) -
(\sqrt{-\gamma}\gamma^{ab})g({\cal D}_a n^{(i)},{\cal D}_b\delta)\,.\cr}
\eqno(3.7)$$
We examine each term separately. For the first term,
we rewrite the argument of ${\cal D}_a$
$$\sqrt{-\gamma}\gamma^{ab} g(n^{(i)}, {\cal D}_\delta e_b) =
{\cal D}_b \left[ \sqrt{-\gamma}\gamma^{ab} g (n^{(i)},\delta)\right] -
\sqrt{-\gamma}\gamma^{ab} g({\cal D}_b n^{(i)},\delta)
-{\cal D}_b(\sqrt{\gamma}\gamma^{ab})\Phi^{(i)} \,,$$
so that
$${\cal D}_a \left[\sqrt{-\gamma}\gamma^{ab}
g(n^{(i)}, {\cal D}_\delta e_b)\right] =
\sqrt{-\gamma}\left[ \Delta \Phi^{(i)} - \nabla_a (T^{a\,(i)(j)}
\Phi^{(j)})\right]\,,$$
where we have introduced the surface torsion
$$T^{(i)(j)}_a = g({\cal D}_a n^{(i)}, n^{(j)})= -T^{(j)(i)}_a\,.\eqno(3.8)$$
The second term gives
$$\eqalign{{\cal D}_\delta (\sqrt{-\gamma}\gamma^{ab})  g({\cal D}_a n^{(i)},
e_b)
=& -{\cal D}_\delta (\sqrt{-\gamma}\gamma^{ab})  K^{(i)}_{ab}\cr
=&  \sqrt{-\gamma} {\cal D}_\delta (\gamma_{ab}) K^{(i)\,ab}\,.\cr}$$
In the last line we use the background equations of motion Eq.(2.7)
to justify dropping a term proportional to $K^{(i)}$.
We can decompose
$$\eqalign{{\cal D}_\delta (\gamma_{ab})= &
g({\cal D}_\delta e_a, e_b) + g(e_a, {\cal D}_\delta e_b)\cr
=& 2 K^{(j)}_{ab} \Phi^{(j)}\,.\cr}$$
We could have used this result directly to derive the equations of motion,
Eq.(2.7).

To evaluate the third term, we note that
$$g({\cal D}_a n^{(i)}, {\cal D}_b\delta) =
T^{(i)(j)}{}_{a} \nabla_b\Phi^{(j)}
+ D_\mu n^{(i)\,\nu} D_\alpha n^{(j)\,\nu} \Phi^{(j)}\,,$$
so that
$$\eqalign{(\sqrt{-\gamma}\gamma^{ab})g({\cal D}_a n^{(i)},{\cal D}_b\delta)
=&\sqrt{-\gamma}\left[\gamma^{ab}T^{(i)(j)}{}_{a} \nabla_b\Phi^{(j)}
+
h^{\mu\nu}D_\mu n^{(i)\,\alpha}{} D_\nu n^{(j)}_{\alpha}\Phi^{(j)}\right]\cr
=&\sqrt{-\gamma}\left[\gamma^{ab}T^{(i)(j)}_{a} \nabla_b
+  K^{(i)}_{ab} K^{(j)\,ab} -
T^{(i)(k)\, a} T^{(k)(j)}_a\right]\Phi^{(j)}\,,\cr}$$
using the definition (3.6) of the projection tensor.

We now add the three terms on the RHS of Eq.(3.7).
The action is given by

$$\eqalign{S=&{1\over2}\int d^D\xi \sqrt{-\gamma}\Bigg(\Phi^{(i)}
\Delta \Phi^{(i)} -2\Phi^{(i)}T^{(i)(j) a}\nabla_a\Phi^{(j)}
+ \Phi^{(i)}\nabla_a T^{(i)(j) a}\Phi^{(j)}\cr
&\quad\quad + \Phi^{(i)}\left[^N\! R_{\mu\nu} n^{(i)\,\mu} n^{(j)\,\nu}
- ^N\! R_{\mu\alpha\nu\beta}\, n^{(i)\,\mu} n^{(k)\,\alpha} n^{(j)\,\nu}
n^{(k)\,\beta}\right]\Phi^{(j)}\cr
&\quad\quad +\Phi^{(i)}\left[K^{(i)}_{ab} K^{(j)\,ab}
+ T^{(i)(k)\, a} T^{(k)(j)}_a\right]\Phi^{(j)}\Bigg)\,.\cr}\eqno(3.9)$$
This coincides with Eq.(A.6) of the third paper in Ref.[2] when
the background geometry is Minkowski space.
The term involving the world-sheet divergence of $T^{(i)(j)}_a$
can be dropped
because it involves a contraction on the normal indices of a term
which is symmetric with a term which is anti-symmetric in these indices.
Such a term will however show up in the equations of motion.
Let us now define

$$\tilde\nabla_a^{(i)(j)}=\nabla_a \delta^{(i)(j)} - T_a^{(i)(j)},.
\eqno(3.10)$$
Then
$$S={1\over2}\int d^D\xi \sqrt{-\gamma}\Bigg(\Phi^{(i)}
\tilde\Delta^{(i)(j)} \Phi^{(j)} -\Phi^{(i)}(M^2)_{(i)(j)}\Phi^{(j)}\,,
\Bigg)\eqno(3.11)$$
where
$$\tilde\Delta^{(i)(j)} =\tilde\nabla^{a(i)(k)}
\tilde\nabla_a^{(k)(j)}\,,\eqno(3.12)$$
and
$$(M^2)_{(i)(j)}=
^N\!R_{\mu\alpha\nu\beta} n^{(i)\,\mu} n^{(k)\,\alpha} n^{(j)\,\nu}
n^{(k)\,\beta}-^N\! R_{\mu\nu} n^{(i)\,\mu} n^{(j)\,\nu}
-K^{(i)}_{ab} K^{(j)\,ab}\,.\eqno(3.13)$$
All three terms involving torsion get absorbed into the
definition of $\tilde\Delta^{(i)(j)}$.

\vskip1pc
\noindent{\bf IV. THE LINEARIZED EQUATIONS}
\vskip2pc

The variation of Eq.(3.9) with respect to $\Phi^{(i)}$ gives
$$\tilde \Delta^{(i)(j)} \Phi^{(j)} -
(M^2)_{(i)(j)}\Phi^{(j)}=0\,. \eqno(4.1)$$
This is a system of $N-D$ non-trivially coupled
scalar wave equations for the $\Phi^{(i)}$ on the curved background
geometry of the world-sheet.

It is worthwhile at this point to comment on the
geometric role played by torsion
in Eq.(4.1).  The sum of the three torsion
terms (the derivative coupling, the
divergence of $T^{(i)(j)a}$ and the quadratic in $T^{(i)(j)a}$)
appearing here render Eq.(4.1)
covariant under local normal frame rotations.

Under a local normal frame rotation $O^{(i)(j)}(\xi)$

$$n^{(i)}\to O^{(i)(j)}(\xi)n ^{(j)}\,,\eqno(4.2)$$
we note that
$\tilde\nabla_a^{(i)(j)}\Phi^{(j)}$ transforms covariantly. This is
because the torsion transforms like a vector potential:

$$T^{(i)(j)}_a \to O^{(i)(k)}T^{(k)(l)}_a (O^{-1})^{(l)(j)} +
({\cal D}_a O O^{-1})^{(i)(j)}\,.\eqno(4.3)$$
The torsion is not itself gauge invariant.
The gauge invariant measure of the torsion is its
curvature defined by

$$\left[\tilde\nabla_a^{(i)(k)}\tilde\nabla_b^{(k)(j)}-
\tilde\nabla_b^{(i)(k)}\tilde\nabla_a^{(k)(j)}\right]\Phi^{(j)}=
T^{(i)(j)}{}_{ab}\Phi^{(j)}\,,$$
so that
$$T^{(i)(j)}{}_{ab}=\nabla_b T^{(i)(j)}_a - T^{(i)(k)}_a T^{(k)(j)}_b
- (a\leftrightarrow b)\,.\eqno(4.4)$$
A choice of normals such that the torsion vanishes exists if and
only if $T^{(i)(j)}{}_{ab}=0$.
However, if one decides to be perverse with one's choice of normals
one can always introduce a torsion.

It is always possible to orient the normals along  a curve
$\xi^a=\Xi^a(s)$ in $m$ such that the torsion vanishes along that curve.
This can be accomplished
by re-orienting the normals at parameter $s$ with the rotation matrix

$$ O(s)^{(i)(j)} = O(0)^{(i)(k)}
P \left(\exp [-\int_0^s ds^\prime \dot \Xi^a (s^\prime) T_a (\Xi(s^\prime))]
\right)^{(k)(j)}\,,$$
where $P$ represents the path ordered product.
This is an analog of the well known result that a coordinate
system always exists in which the Riemannian connection
vanishes along any given curve.

There are two ways that the scalar fields can couple.
One way is through the effective
mass matrix $(M^2)_{(i)(j)}$ given by Eq.(3.13) If there is
torsion, however, they can also couple though
$\tilde\Delta$.
Though $T^{(i)(k)\, a} T^{(k)(j)}_a$ couples the $\Phi^{(i)}$
like a mass term, it is more naturally grouped in the
combination appearing in the definition of $\tilde\nabla$.
$(M^2)_{(i)(j)}$ need not possess a
global sign. It can be diagonalized at
any point with its eigenvalues forming its diagonal entries.
If the world-sheet were Minkowski space,
a negative eigenvalue of $M^2_{(i)(j)}$ would
signal an instability. However, in general,
there is no simple correlation between tachyonic masses and instabilities.
An explicit counter-example is provided by
perturbation theory about a class of defects in de Sitter space
discussed in Ref.[2] and which we will examine below.

In the case of a domain wall with a single normal vector, both the
torsion and the total projection of the spacetime Riemann curvature
onto the normal vanish. Eq.(4.1) then reduces to the form
$$\Delta \Phi + \left[R_{\mu\nu} n^{\mu} n^{\nu}
+K^{ab} K_{ab}\right]\Phi=0\,.\eqno(4.1^\prime)$$
In this case, the quadratic in the
extrinsic curvature can be eliminated in favor of intrinsic
geometric scalars using the contracted Gauss-Codazzi equations.
We reproduce Eq.(4.1) of Ref.[2] with $\rho=0$.
Henceforth, we will assume that the
the co-dimension of the defect exceeds one.

Even when the background geometry is flat so that
$^N\! R^\mu{}_{\nu\alpha\beta}=0$,
Eq.(4.1) is extremely complicated, involving scalars in the
extrinsic geometry
($K^{(i)}_{ab}$ and $T^{(i)(j)}_a$) in combinations which, it appears,
cannot be eliminated in favor of intrinsic geometric scalars.
To see this, let us recall the equations for
the embedding.[5] These are
$$g(^N\! R(e_a,e_b)e_c,e_d) = ^D\!R_{abcd} +
K_{ac}^{(i)} K_{bd}^{(i)} - K^{(i)}_{ad} K^{(i)}_{bc}
\,,\eqno(Gauss-Codazzi)$$
$$g(^N\! R(e_a,e_b)e_c, n^{(i)}) = \nabla_a K_{bc}^{(i)} +
T^{(i)(j)}_b K_{ac}^{(j)} - (a\leftrightarrow b)\,,\eqno(Codazzi-Mainardi)$$
and
$$g(^N\! R(e_a,e_b)n^{(i)},n^{(j)}) =
T^{(i)(j)}{}_{ab} -
\left[K^{(i)}_{ac} K^{(j)\,c}_b
- (a\leftrightarrow b)\right]\,,\eqno(Ricci)$$
where $T^{(i)(j)}{}_{ab}$ is given by Eq.(4.4).
Note that the torsion only occurs in
gauge invariant combinations in these equations.

Unlike the case of a hypersurface, we cannot exploit the
Gauss-Codazzi equations to eliminate
the quadratic in $K_{ab}^{(i)}$ in favor of spacetime and
world-sheet curvature scalars.
This is because it is the traced product
over the normal indices, $K^{(i)}_{ab}K^{(i)}_{cd}$
which appears in these equations.
The quadratics in $T^{(i)(j)}_a$ which appear in the Ricci equations
are anti-symmetric in both world-sheet and normal indices. These
equation therefore do not help us to eliminate the quadratic in
$T^{(i)(j)}_a$ appearing in Eq.(4.1).

If, however, we can choose our normal vectors such that
all but one of them, say $n^{(1)}$, are parallel transported
along any curve on the world-sheet
$${\cal D}_a n^{(i)}=0\,,\eqno(4.5)$$
then $T^{(i)(j)}_a=0$ for all $i$ and $j$.
The vanishing of $T^{(1)(j)}_a$ is assured by
the anti-symmetry of $T^{(i)(j)}_a$ with respect to its
normal indices.	We thus identify a sufficient set
of conditions on the embedding under which the surface torsion vanishes.

In addition, the conditions Eq.(4.5) imply that the only
linear combination of extrinsic curvature tensors which is
non-vanishing is the one that corresponds to the exceptional
normal direction:
$$K^{(i)}_{ab}=0,\quad i=2,\cdots N-D\,.\eqno(4.6)$$
Only one of the
background equations of motion will now be
non-trivial. The marvellous thing about Eq.(4.5) for our
present puposes is that
the quadratic in $K_{ab}^{(i)}$ appearing in Eq.(4.1)
can be replaced by its trace:
$$K^{(i)}_{ab} K^{(j)\,ab} =
\delta ^{(i)(1)}\delta^{(j)(1)}K^{(k)}_{ab} K^{(k)\,ab} \,,$$
so that the contracted Gauss-Codazzi equation can be used
exactly as it was in the case of a hypersurface to
eliminate it in terms of curvatures:
$$\eqalign{K^{ab\,(i)} K_{ab}^{(j)} =&
^N\! R_{\mu\nu\alpha\beta} h^{\mu\alpha} h^{\nu\beta} -^D\!R\cr
=&^N\! R - 2^N\! R_{\mu\nu} n^{(k)\,\mu}n^{(k)\,\nu} +
2^N\! R_{\mu\nu\alpha\beta}\,
n^{(k)\,\mu} n^{(k)\,\alpha} n^{(\ell)\,\nu} n^{(\ell)\,\beta}
 -^D\!R\,,\cr}\eqno(4.7)$$
where we have exploited the definition of the projection tensor,
Eq.(3.8). The normal projections of the spacetime Ricci and Riemann tensors
are of a kind already encountered in Eq.(4.1). However, here they
do not imply any coupling between different $\Phi^{(i)}$'s.

To provide a geometrical picture for what Eq.(4.5) implies, it is
useful to recall the form of the
Gauss-Weingarten equations [5] which describe the change in the
basis vectors as one moves about the surface
$$\eqalign{ {\cal D}_a e_b =&\gamma^c_{ab} e_c + K^{(i)}_{ab} n^{(i)}\cr
	    {\cal D}_a n^{(i)} =& -K^{(i)}_{ab} e^b + T^{(i)(j)}_a n^{(j)}\,,\cr}
\eqno(4.8a,b)$$
where the $\gamma^c_{ab}$ are the worldsheet connection coefficents.
The embedding equations are the integrability conditions
ensuring that a solution to these equations exists.
When Eq.(4.5) is satified,
only $n^{(1)}$ changes as we
move about the worldsheet and Eqs.(4.8)
reduce to the form
$$\eqalign{ {\cal D}_a e_b =&\gamma^c_{ab} e_c + K^{(1)}_{ab} n^{(1)}\cr
	    {\cal D}_a n^{(1)} =& -K^{(1)}_{ab} e^b\,, \cr}
\eqno(4.8^\prime)$$
and
$${\cal D}_a n^{(I)} =0\,, \quad I=2,\cdots, N-D\,.$$
The first two equations are simply the hypersurface form of the
Gauss-Weingarten equations. The world-sheet can be
embedded as a hypersurface in a $D+1$
dimensional sub-manifold of $M$, let us say ${\cal M}$.

Let us now look for conditions on the
geometry in the neighborhood of the world-sheet making it
consistent with Eq.(4.5).
To do this, we construct a
coordinate system for $M$ adapted to ${\cal M}$
in the neighborhood of $m$.
Let $y^A$, $A=0,\cdots, D$ be coordinates for
${\cal M}$ in this neighborhood.
We now complete the
coordinate system for $M$ by complementing the coordinates on ${\cal M}$
with $N-D-1$ coordinates $\{z^I\}$, $I=2,\cdots,N-D$ such that
${\cal M}$ is given by $z^I=0$. The normals to $m$,
$n^{(I)}$, $I=2,\cdots,N-D$ are then linear combinations of the
gradients of the $z^I$ evaluated on $m$.
With respect to these coordinates,
Eqs.(4.5) can be replaced by the following conditions on the
spacetime metric evaluated on $m$:

$$\eqalign{ g_{AB,I}=&0\cr
	    g_{AI}=&0\,.\cr}\eqno(4.9)$$
Any defect in Minkowski space which lies in a $D$ dimensional
plane will satisfy these conditions.

Let us describe de Sitter space
by a FRW closed line element

$$ds^2=-dt^2 + H^{-2} \cosh ^2 (Ht) d\Omega^2_{N-1}\,,$$
where
$d\Omega^2_{N}$ is the line element on a round $N-1$ sphere and
$H$ is the Hubble parameter.
The subspace consisting of any number of fixed azimuthal angles
is also a de Sitter space with the same Hubble parameter.
A  $D$ dimensional defect in
de Sitter space with  $N-D-1$  fixed
azimuthal angles will also satisfy Eq.(4.9).

In the two cases considered above, spacetime is
homogeneous and isotropic.
A less trivial example satisfying Eq.(4.9)
is a string in Schwarzschild space on a
fixed value of the azimuthal angle.
Thus we see that Eqs.(4.9) is consistent with
a reasonably rich class of geometries.

Let us suppose, in addition, that the geometry of the world sheet
is spherically symmetric, {\it i.e}
invariant under the rotation group, $O(D-1)$. The world-sheet of the
defect is then a $D$ dimensional FRW homogeneous and
isotropic closed universe described by the
line element
$$ds^2 =-d\tau^2 + a^2(\tau) d\Omega^2_{D-1}\,,$$
where $\tau$ is the proper time registered on a co-moving clock.
The function $a(\tau)$ is the proper
circumferential radius $r$ on the $D-1$ sphere at proper time $\tau$.
Consistency then demands that the spacetime metric be
invariant under some $O(d-1)$ with $d\ge D$
with $D-1$ common axes of symmetry.
The only non-trivial dynamics now takes place in a
$1+1$ dimensional subspace of ${\cal M}$.
The Gauss-Weingarten equations mimic the
Frenet-Serret equations describing the motion
of a particle in this two-dimensional spacetime:

$$\eqalign{ {\cal D}_\tau e_\tau =& K^{(1)}_{\tau\tau} n^{(1)}\cr
	    {\cal D}_\tau n^{(1)} =& -K^{(1)}_{\tau\tau} e^\tau \,.\cr}
\eqno(4.10)$$
The condition $\gamma^a_{\tau\tau} =0$ is the analog of the
statement the acceleration along a timelike
curve is orthogonal to the velocity when the
curve is pamametrized by proper time.

There is another simplification which occurs whenever $D=N-2$,
an example of which is provided by a string in any four dimensional
manifold. For then, the coupling between the two
scalar field components
$\Phi^{(1)}$ and $\Phi^{(2)}$ which is mediated by the terms of the
form

$$^N\! R_{\mu\alpha\nu\beta}\, n^{(i)\,\mu} n^{(k)\,\alpha} n^{(j)\,\nu}
n^{(k)\,\beta}\Phi^{(j)}\eqno(4.11)$$
in Eq.(4.1) vanishes. Let $i=1$. Then, the only surviving term in Eq.(3.13)
is

$$^N\! R_{\mu\alpha\nu\beta} n^{(1)\,\mu} n^{(2)\,\alpha} n^{(1)\,\nu}
n^{(2)\,\beta}\,\Phi^{(2)}$$
--- the fields decouple. This condition may not be
independent of Eqs.(4.5).

If the background spactime is Einstein, with cosmological
constant $\Lambda$,
$$R_{\alpha\beta}={2\Lambda\over N-2} g_{\alpha\beta}\,,$$
the Ricci curvature coupling between different scalar fields
disappears:
$$R_{\mu\nu} n^{(i)\,\mu} n^{(j)\,\nu} ={2\Lambda\over N-2}\delta^{(i)(j)}\,.$$
If the background is de Sitter space,
the Riemann curvature coupling also disappears independent of the
dimension of the defect.
For then
$$R_{\mu\alpha\nu\beta}=H^2\left[
g_{\mu\nu}g_{\alpha\beta}-
g_{\mu\beta} g_{\nu\alpha}\right]\,,\eqno(4.12)$$
so that
$$R_{\mu\alpha\nu\beta} n^{(i)\,\mu} n^{(k)\,\alpha} n^{(j)\,\nu}
n^{(k)\,\beta} =
{H^2} (N-D-1)\delta^{(i)(j)}\,.$$

In particular, in the case of a
any defect lying on the subspace
with $N-D-1$ fixed aximuths ( it need not itself be
spherically symmetric) in de Sitter space,
the best of all worlds is realized.
The scalar fields completely decouple. Without any essential loss of
generality, we will consider co-dimension two. Now

$$\eqalign{-(M^2)_{(1)(1)} =&
^N\!R + ^N\! R_{\mu\nu} n^{(1)\,\mu} n^{(1)\,\nu}
-^N\! R_{\mu\nu} n^{(k)\,\mu} n^{(k)\,\nu}\cr
&+^N\! R_{\mu\alpha\nu\beta}\, n^{(1)\,\mu} n^{(2)\,\alpha} n^{(1)\,\nu}
n^{(2)\,\beta} -^{N-2}\! R\,.\cr}\eqno(4.13a)$$
and
$$-(M^2)_{(2)(2)}=
 ^N\! R_{\mu\nu} n^{(2)\,\mu} n^{(2)\,\nu}
- ^N\! R_{\mu\alpha\nu\beta}\, n^{(1)\,\mu} n^{(2)\,\alpha} n^{(1)\,\nu}
n^{(2)\,\beta}\,.\eqno(4.13b)$$
We substitute Eq.(4.12) into Eqs.(4.13) to get
$$\eqalign{-(M^2)_{(1)(1)}=&(N-2)H^2+
^D\!R - (N-2)(N-3) H^2 ,,\cr
-(M^2)_{(2)(2)}=&(N-2)H^2\,.\cr}\eqno(4.14)$$
$M_{(2)(2)}$ is independent of the motion of the defect and is always
tachyonic. Both wave equations  depend only
on the intrinsic geometry of the worldsheet.

Let us now specialize to spherically symmetric defects of
co-dimension two. In four dimensions these are circular strings.
A circular string string can follow two qualitatively different
trajectories $a=a(\tau)$ in de Sitter space.[1]
One of these consists of trajectories
which begin with $a=0$ grow to a
maximum value before recollapsing to $a=0$. The other
is the bounce which consists of a trajectory
originating on the equator contracting to a minimum
value and then bouncing back to the
equator. The discription in terms of $a(\tau)$ is qualitatively
identical to that for spherical domain walls.

In particular, there is a bounce which does not does really bounce
at all representing a circular string which spans the equator.
This solution can be interpreted as a string which tunnels from nothing
due to quantum mechanical processes. [1]
These are the strings about which perturbation theory was
examined in Ref.[2].

The world-sheet is now an embedded
$N-2$ dimensional de Sitter space
with
$$^{N-2}\! R = (N-2)(N-3) H^2\,.\eqno(4.15)$$
We note that now, not only do Eqs(4.5) hold, but in addition
${\cal D}_a n^{(1)}=0$.
The normal directions $n^{(1)}$ and $n^{(2)}$ are now
entirely equivalent.  Our construction.
which did not  exploit the
extra symmetry of a defect which spans the equator degenerates.
Geometrically, $K^{(i)}_{ab}=0$ for all $i$. In
mathematical parlance, the world-sheet is totally geodesic [5].

It should not be surprising that
perturbation theory simplifies dramatically in this case.
The two effective mass eigenvalues now coincide and are tachyonic.
The equations of motion for $\Phi^{(1)}$ and $\Phi^{(2)}$ are
therefore identical:
$$\Delta\Phi^{(1),(2)} +(N-2)H^2\Phi^{(1),(2)}=0\,,\eqno(4.16)$$
reproducing the expression obtained in Ref [2].
The wave equation for each component
mimics the equation for
an equatorial domain wall in an $N-1$ dimensional de Sitter spacetime.
We note that the technique used in ref.[2] to derive
Eq.(4.14) depended sensitively
on the fact that the embedded domain wall spanned the
equator. Now, however, we possess a general framework
which not only has permitted us to predict that decoupling would occur but
also explains why the effective masses coincide in this geometry.

\vskip1pc
\centerline{\bf V. CONCLUSIONS}
\vskip1pc
We have provided a framework for the examination of
perturbations on topological defects on a given spacetime background
which generalizes our earlier work on domain walls
to lower dimensional topological defects.
In either case, however,
the coupling of the perturbation to extrinsic geometry makes the theory
very different from the scalar field theories we are familiar with.
When the co-dimension of the world-sheet is $r$, there will be
$r$ scalar fields describing the perturbation. What is more, the
equations we obtain are generally coupled in a non-trivial way.
There is a coupling through an effective mass matrix involving
quadratics in the extrinsic curvature as well as appropriate
projections of the spacetime Riemann curvature. In addition, however,
on a lower dimensiuonal defect
there will be a coupling between the scalar fields  mediated
by the torsion of the embedding. We have
examined the geometrical role played by torsion in the formalism.
On one hand this coupling ensures that the
equations of motion transform covariantly
under local normal frame rotations. As
such, it plays the role of a vector potential coupling to the
scalar field. The only invariant measure of the
tosion is its curvature. If the
curvature vanishes the torsion can be gauged away by an appropriate
local rotation of the normal vectors.

If the geometry under consideration
possesses a symmetry some simplification is always likely.
We focused on the identification of a sufficient set of
simpifying conditions without any attempt to be rigorous. We showed,
however,  that these conditions
are realized under geometrical conditions
which are sufficiently general to be useful.
When, in particular, the background is de Sitter space and the
defect is oriented along any number of fixed azimuthal angles,
the scalar fields completely decouple.

A challenge is to formulate a consistent quantum field theory
of perturbations. The renormalization of the theory will require
the addition of counterterms involving extrinsic geometry.

The formalism should also prove useful for the
examination of fluctuations about instantons in
the semi-classical approximation to tunneling.
Now the signature of both the background spacetime and the world-sheet
is Euclidean. One is then interested in the eigen-modes of the
corresponding Euclidean operator
$$(\tilde \Delta \Phi)^{(i)} -(M^2)_{(i)(j)}\Phi^{(j)}=\lambda\Phi^{(i)}\,,$$
in particular those which correspond to negative or zero eigenvalues.

The weak point in our analysis is that it fails to treat the
topological defect as a source for gravity. We are currently
addressing this problem in the context of domain walls. The
treatment of the perturbation is, however,
likely to be problematical for co-dimensions higher than one. As such,
it would probably be more rewarding to
examine perturbation theory
in the context of a field theoretical model for the
topological defect.

\vskip2pc
\centerline{\bf ACKNOWLEDGEMENTS}
\vskip1pc
It is a pleasure to acknowledge valuable conversations with
Jaume Garriga and Alexander Vilenkin. I am
particularly grateful to them for pointing out a blunder
in my initial calculation. After this work was completed I received a
preprint from Arne Larsen (hep-th/9303001)
who also discusses perturbation theory about strings from a
covariant point of view.

\vfill
\eject
\centerline{\bf REFERENCES}
\vskip1pc
\item{1.} R. Basu, A.H. Guth and A. Vilenkin
{\it Phys Rev} {\bf D44} (1991) 340
\vskip1pc
\item{2.} J. Garriga and A. Vilenkin {\it Phys. Rev} {\bf D44} (1991) 1007;
{\it Phys. Rev} {\bf D45} (1992) 3469 and
{\it Phys. Rev.} {\bf D47} (1993) 3265 In the appendix to the
third paper, perturbation theory is developed
about an arbitrary defect in Minkowski space.
\vskip1pc
\item{3.} J. Guven {\it ICN Preprint} `` Covariant Perturbations of
Domain walls''(1993)
\vskip1pc
\item{4.} S.W. Hawking and G.F.R. Ellis {\it The Large Scale Structure of
Space-Time} (Cambridge Univ. Press, Cambridge, 1973)
\vskip1pc
\item{5.} The classical text is
Eisenhart {\it Riemannian Geometry} (Princeton Univ. Press,
Princeton, 1947); For a more modern point of view
M. Spivak {\it Introduction to Differential Geometry} Vol. IV
(Publish or Perish, Boston MA);
Hypersurface embeddings are discussed in
K. Kucha\v r {\it J. Math. Phys} {\bf 17} (1976) 777,792,801

\vskip1pc
\item{6.} M. Green, J. Schwarz and E. Witten {\it Superstring Theory}
(Cambridge, 1987)
\vskip1pc
\item{7.} For a review of the subject in the context of Harmonic maps,
see J. Eels and L. Lemaire {\it Bull. London Math Soc.}
{\bf 10} (1978) 1
\vskip1pc
\item{8.} One could, however, always construct
a system of coordinates adapted to the worldsheet in the manner of
ADM. See, for example, C. Misner, K. Thorne and J.A.
Wheeler {\it Gravitation} (Freeman, San Francisco
1974). This would involve the introduction of matrix
generalizations of the lapse and shift.

\bye